\documentclass[aps,pra,showpacs,superscriptaddress,nofootinbib]{revtex4}
\usepackage{bbm}
\usepackage{mathrsfs}
\usepackage{epsfig,psfrag}
\usepackage{amsmath,amsfonts,amssymb}
\usepackage[usenames]{color}
\usepackage{bm}
\usepackage{ulem}
\definecolor{MidnightBlue}{cmyk}{0.98,0.13,0,0.43}
\definecolor{DarkGreen}{rgb}{0,0.7,0.1}

\newcommand{\sgn}{{\rm sgn}}

\begin{document}

\title{Electromagnetic Casimir Forces of Parabolic Cylinder and
Knife--Edge Geometries}


\author{Noah Graham}
\email{ngraham@middlebury.edu}
\affiliation{Department of Physics,
Middlebury College,
Middlebury, VT 05753, USA}

\author{Alexander Shpunt}
\affiliation{Department of Physics, Massachusetts Institute of
Technology, Cambridge, MA 02139, USA}

\author{Thorsten Emig}
\affiliation{Laboratoire de Physique Th\'eorique et Mod\`eles Statistiques,
CNRS UMR 8626, B\^at.~100,
Universit\'e Paris-Sud,
91405 Orsay cedex,
France}

\author{Sahand Jamal Rahi}
\affiliation{Department of Physics, Massachusetts Institute of
Technology, Cambridge, MA 02139, USA}
\affiliation{Center for Studies in Physics and Biology,
The Rockefeller University,
1230 York Street, New York, NY 10065, USA}

\author{Robert L. Jaffe}
\affiliation{Department of Physics, Massachusetts Institute of
Technology, Cambridge, MA 02139, USA}
\affiliation{Center for Theoretical Physics and Laboratory for Nuclear
Science, Massachusetts Institute of
Technology, Cambridge, MA 02139, USA}

\author{Mehran Kardar}
\affiliation{Department of Physics, Massachusetts Institute of
Technology, Cambridge, MA 02139, USA}

\begin{abstract}

An exact calculation of electromagnetic scattering from a perfectly
conducting parabolic cylinder is employed to compute Casimir forces in
several configurations.  These include interactions between a
parabolic cylinder and a plane, two parabolic cylinders, and a
parabolic cylinder and an ordinary cylinder. To elucidate the effect
of boundaries, special attention is focused on the ``knife-edge''
limit in which the parabolic cylinder becomes a half-plane.  
Geometrical effects are illustrated by considering arbitrary rotations
of a parabolic cylinder around its focal axis, and arbitrary
translations perpendicular to this axis.   A quite different
geometrical arrangement is explored for the case of an ordinary
cylinder placed {\em in the interior} of a parabolic cylinder.  All of
these results extend simply to nonzero temperatures.

\end{abstract}

\maketitle

\section{Introduction}

The Casimir force, arising from quantum fluctuations of the
electromagnetic field in vacuum, is a striking manifestation of
quantum field theory at the mesoscopic scale.  
Casimir's computation of the force between two parallel metallic
plates \cite{Casimir48} gives the classic demonstration of this
phenomenon.  Following its experimental confirmation in the past decade
\cite{experiments}, however, the Casimir force is now important to the
design of microelectromechanical systems \cite{MEMS}.  Potential
practical applications have motivated the development of large-scale
numerical methods to compute Casimir forces for objects of any shape
\cite{Johnson,worldline,Maggs}.  In contrast, the simplest and most
commonly used analytic methods for dealing with complex shapes, such
as the proximity force approximation (PFA), rely on pairwise
summations, limiting their applicability.

Recently we have developed a formalism \cite{spheres,universal} 
that relates the Casimir interaction among several
objects to the scattering of the electromagnetic field from the
objects individually.  This method decomposes the 
path integral representation of the Casimir energy \cite{GK} as a
log-determinant \cite{Klich} in terms of a multiple
scattering expansion, as was done for asymptotic separations in
Ref.~\cite{Balian}.  It can also be regarded as a
concrete implementation of the perspective emphasized by Schwinger
\cite{Schwinger75} that the fluctuations of the electromagnetic field
can be traced back to charge and current fluctuations on the
objects.  (For additional perspectives on the scattering formalism,
see also references in \cite{universal}.)  This approach
allows us to take advantage of the well-developed machinery of
scattering theory.  In particular, the availability of 
exact scattering amplitudes for simple objects, such as spheres and
cylinders, has made it possible to compute the Casimir force for
two spheres \cite{spheres}, a sphere and a plate \cite{sphere+plate},
multiple cylinders \cite{cylinders}, and cases with more than two
objects \cite{JamalAlejandro,Maghrebi}.  This formalism has also been
applied and extended in a number of other situations
\cite{Johnson,Kenneth,Milton,Golestanian,Ttira}.

Here we expand on recent work \cite{paraboloid} that showed how to
apply these techniques to \emph{parabolic} cylinders, another example
where the scattering amplitudes can be computed exactly.  The limiting
case when the radius of curvature at the tip vanishes, so that the
parabolic cylinder becomes a semi-infinite plate (a ``knife-edge''), 
provides a particularly interesting application of this approach.  One
can also model the knife-edge as the limit of a wedge of zero opening
angle \cite{sharp}; the two approaches are rather complementary
as the former is most amenable to {\em numerical} computation, 
while the latter yields approximate {\em analytic} formulae via a multiple 
reflection expansion (which is useful for other sharp geometries, 
such as the cone \cite{sharp}).  Edge geometries have also been
considered in Refs.~\cite{Gies,Kabat}.

The remainder of the manuscript is organized as follows:
The Helmholtz equation in the parabolic cylinder coordinate system is reviewed
in Sec.~\ref{scatter}, and exact formulae are derived for the scattering
of the electromagnetic field from a perfectly conducting parabolic cylinder.
The techniques of Refs.~\cite{spheres,universal}, are then employed to
find the electromagnetic Casimir interaction energy for a variety of
situations: We let the parabolic cylinder interact with a plane
(Sec.~\ref{Pcyl-plane}), a second parabolic cylinder
(Sec.~\ref{2Pcyls}), or an ordinary cylinder (Sec.~\ref{Pcyl-cyl}).  
In these calculations we consider arbitrary
rotations of the parabolic cylinder around its focal axis and arbitrary
translations perpendicular to that axis, transformations that are
particularly useful when considering the ``knife-edge'' limit.  
We also position an ordinary or parabolic cylinder inside a parabolic
cylinder (Sec.~\ref{interior}), and incorporate the effects of thermal
corrections in all of these calculations (Sec.~\ref{finiteT}).

\section{Scattering in Parabolic Cylinder Coordinates}
\label{scatter}

\begin{figure}[htbp]
\includegraphics[width=0.45\linewidth]{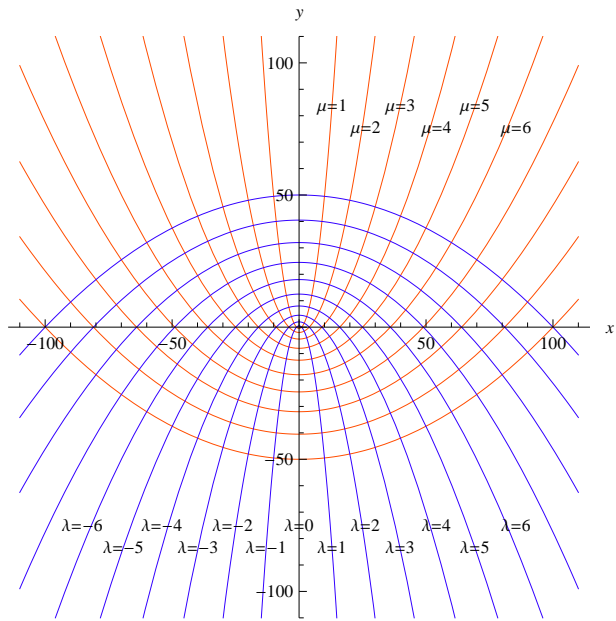}
\caption{Coordinate curves in parabolic cylinder coordinates.}
\label{fig:coords}
\end{figure}

We begin with a review of scattering theory in parabolic cylinder
coordinates \cite{MF,WW}.  Because the system is translationally
invariant in the $z$ direction and perfectly reflecting, we can
decompose the electromagnetic scattering problem into two scalar
problems, one with Dirichlet and the other with Neumann boundary
conditions.  Parabolic cylinder coordinates are defined by
\begin{eqnarray}
x = \mu \lambda\,, \qquad
y = \frac{1}{2} (\lambda^2 - \mu^2) \,\qquad
z = z \,,
\end{eqnarray} 
such that for $r=\sqrt{x^2 + y^2}$,
\begin{equation}
\lambda = \sgn(x) \sqrt{r+y} \,\qquad \mu = \sqrt{r-y} \,,
\end{equation}
where we have chosen a convention where $\mu \geq 0$ but $\lambda$
can have either sign, which is explained in more detail below.  This
restricted domain is sufficient to include all points in space.  Note
that here we must take $\sgn(0) = 1$, not $\sgn(0) = 0$.
At fixed $z$, the surfaces of constant $\mu$ are confocal parabolas
opening upward (toward positive $y$) and the surfaces of constant
$\lambda$ are confocal parabolas opening downward (toward negative
$y$), as shown in Fig.~\ref{fig:coords}.  The scale factors (metric
coefficients) are $h_\lambda = h_\mu = \sqrt{\lambda^2 + \mu^2}$,
$h_z = 1$. 

We would like to solve the Helmholtz equation, which in these
coordinates takes the form
\begin{equation}
\nabla^2 \Phi(\bm{r}) = \frac{1}{\lambda^2 + \mu^2} 
\left(\frac{d^2 \Phi}{d\lambda^2} + \frac{d^2 \Phi}{d\mu^2} \right) 
+ \frac{d^2 \Phi}{dz^2} = -k^2 \Phi(\bm{r}) \,,
\end{equation}
where eventually we will set $k = i \kappa$.  For $k$ real, we expect
oscillating traveling wave solutions, while for $\kappa$ real, we
expect exponentially growing and decaying solutions.  This equation is
amenable to separation of variables:
\begin{equation}
\Phi(\lambda, \mu, z) = L(\lambda)M(\mu)  Z(z)\,.
\end{equation}
Separation of the $z$ variable is trivial,
$Z(z) = e^{ik_z z}$, leaving 
\begin{equation}
\frac{1}{L}\frac{d^2 L}{d\lambda^2} + \frac{1}{M}\frac{d^2 M}{d\mu^2} = 
(k_z^2 - k^2) (\lambda^2 + \mu^2)\,,
\end{equation}
which gives the separated equations
\begin{eqnarray}
\frac{d^2 L}{d\lambda^2} - (k_z^2 -k^2) \lambda^2 L &=& - q L \,, \cr
\frac{d^2 M}{d\mu^2} - (k_z^2 -k^2) \mu^2 M  &=& q M \,,
\end{eqnarray}
where $q$ is a separation constant.  The solutions are
\begin{eqnarray}
L(\lambda) =  D_\nu(\tilde \lambda) &\hbox{and}& 
L(\lambda) = D_{-\nu-1}(i\tilde \lambda) \,, \cr
M(\mu) =  D_\nu(i\tilde \mu)  &\hbox{and}& 
M(\mu) = D_{-\nu-1}(\tilde \mu) \,,
\end{eqnarray}
in terms of the parabolic cylinder function $D_\nu(u)$.  Here 
\begin{eqnarray}
\tilde \lambda &=&  \lambda \sqrt{-2i \sqrt{k^2 - k_z^2}}
= \lambda \sqrt{2\sqrt{k_z^2 + \kappa^2}} \,, \cr
\tilde \mu &=& \mu \sqrt{-2i \sqrt{k^2 - k_z^2}}
=\mu \sqrt{2\sqrt{k_z^2 + \kappa^2}} \,,
\end{eqnarray}
and
\begin{equation}
\nu = \frac{1}{2}\left(\frac{q}{\sqrt{k_z^2 - k^2}}
 - 1\right)\,, \hbox{~~~which implies~~~}
-\nu-1 = \frac{1}{2}\left(\frac{-q}{\sqrt{k_z^2 - k^2}} - 1\right)
\,.
\end{equation}
The second solution in each case is
obtained by making the replacements $q\to -q$, $\lambda\to i\lambda$, and
$\mu\to -i\mu$, the combination of which leaves the differential
equations invariant.
For $\displaystyle |\arg u| < {3\pi}/{4}$,
$D_\nu(u) \sim u^{\nu} e^{-u^2/4}$ as $|u|\to \infty$.  So for both
$\lambda$ and $\mu$, we have solutions that both grow and decay
exponentially as their argument approaches positive
infinity.  Since the Cartesian radial distance is $\displaystyle r = 
(\lambda^2 + \mu^2)/2$, these are ordinary exponentials (not
Gaussians) when expressed in terms of Cartesian coordinates.  

If we send $\lambda$ to $-\lambda$ and $\mu \to -\mu$,
we return to the same point in space, so a
wavefunction defined everywhere in the plane must be
invariant under this transformation.  From the identity
\begin{equation}
D_\nu(-\tilde \lambda) = (-1)^\nu D_\nu(\tilde \lambda) +
\frac{\sqrt{2\pi}}{\Gamma(-\nu)} i^{\nu+1} D_{-\nu-1}(i\tilde \lambda)\,,
\end{equation}
we can conclude that $\nu$ must be a non-negative integer to satisfy this
requirement.  We begin with the case
\begin{equation}
L(\lambda)M(\mu) = D_\nu(\tilde \lambda) D_{\nu}(i\tilde \mu)\,,
\label{eqn:regular}
\end{equation}
for $\nu=0,1,2,3,\cdots$. For these values of $\nu$, the parabolic
cylinder functions with real arguments are simple rescalings of
the solutions to the quantum harmonic oscillator, and thus are given
by a Gaussian times a Hermite polynomial.  The combined solution
in Eq.~(\ref{eqn:regular}) then takes the form of a polynomial in $\mu$ and
$\lambda$ times $\displaystyle e^{i \sqrt{k^2-k_z^2} y}$, and thus
represents a traveling parabolic wave in the $+y$ direction.

These solutions represent ``regular'' waves, the analogs of solutions
in spherical coordinates involving spherical Bessel functions and
spherical harmonics, $j_\ell(kr) Y^{\ell}_m(\theta, \phi)$.  Since we
have chosen to restrict $\mu$ to positive values, it will represent
the analog of the radial coordinate $r$.  We will also require
``outgoing'' solutions to the same differential equations, the analogs of
solutions in spherical coordinates involving spherical Hankel functions and
spherical harmonics, $h^{(1)}_\ell(kr) Y^{\ell}_m(\theta, \phi)$.
As in the spherical case, in the irregular solution
the function of the ``angular'' variable
$\lambda$ is the same, but the function of the ``radial'' variable is
an independent solution to the same differential equation,
\begin{equation}
L(\lambda)M(\mu) = D_\nu(\tilde \lambda) D_{-\nu-1}(\tilde \mu) \,,
\label{eqn:irregular}
\end{equation}
again for $\nu=0,1,2,3,\cdots$.  Even though they do not blow up at
$\mu=0$ (as the outgoing spherical wavefunctions do at $r=0$), these
solutions are not permissible for $\mu < 0$ because they are not
invariant under the combined substitution $\mu \to -\mu$ and $\lambda
\to -\lambda$.  The solution in Eq.~(\ref{eqn:irregular})
asymptotically approaches a polynomial in $\mu$ and $\lambda$ times
$\displaystyle e^{i \sqrt{k^2-k_z^2} r}$, and thus represents an
outgoing radial parabolic wave.

We define the full regular and outgoing solutions
\begin{eqnarray}
\psi_\nu^{\hbox{\tiny reg}}(\bm{r}) &=& 
i^\nu e^{i k_z z} D_\nu(\tilde \lambda)  D_\nu(i\tilde \mu)\,,  \cr
\psi_\nu^{\hbox{\tiny out}}(\bm{r}) &=& 
e^{i k_z z} D_\nu(\tilde \lambda)  D_{-\nu-1}(\tilde \mu) \,,
\end{eqnarray}
using which the free Green's function becomes, for $\mu\geq 0$
\cite{MF},\footnote{The factor of $(-1)^\nu$ is incorrectly omitted
in Ref.~\cite{MF}.}
\begin{equation}
G(\bm{r}_1, \bm{r}_2, k)
= \int_{-\infty}^\infty \frac{d k_z}{2 \pi}
\sum_{\nu=0}^\infty \frac{(-1)^{\nu}}{\nu!\sqrt{2\pi}}
\psi_\nu^{\hbox{\tiny reg}}(\bm{r_<})^*
\psi_\nu^{\hbox{\tiny out}}(\bm{r_>}) \,.
\label{eqn:Green1}
\end{equation}
Here $\bm{r}_<$ ($\bm{r}_>$) is the point with the smaller (larger) of
$\mu_1$ and $\mu_2$, and we have made use of the Wronskian of the two
independent solutions for each $\nu$,
\begin{equation}
W[D_\nu(u),D_{-\nu-1}(i u)] = i^{\nu-1} \,.
\end{equation}

The decomposition of a plane wave in regular parabolic cylinder
functions is \cite{MF}
\begin{equation}
e^{i\bm{k}\cdot\bm{r}}
= e^{i k_z z} \frac{1}{\cos \frac{\phi}{2}}
\sum_{\nu=0}^\infty \frac{1}{\nu!}
\left(\tan \frac{\phi}{2} \right)^{\nu}
\psi_\nu^{\hbox{\tiny reg}}(\bm{r}) \,,
\label{eqn:plane1}
\end{equation}
where
\begin{equation}
\phi = \frac{1}{2i} \log \frac{k_y + ik_x}{k_y - ik_x}\,,
\label{eq:phi}
\end{equation}
and $k_y=\sqrt{k^2-k_x^2-k_z^2} = i\sqrt{\kappa^2+k_x^2+k_z^2}$.
Here the logarithm defines the arctangent of $k_x/k_y$ in the
appropriate quadrant.  Note that the expansion in
Eq.~(\ref{eqn:plane1}) converges only for $k_y>0$, since it is built
out of parabolic waves that propagate upward.

To determine the T-matrix we consider Dirichlet or Neumann boundary
conditions at $\mu=\mu_0 \geq 0$.  In the region $\mu > \mu_0$
we have the scattering solution
\begin{equation}
\Phi(\bm{r}) = 
\psi_\nu^{\hbox{\tiny reg}}(\bm{r})
 - i^\nu \frac{D_{\nu}(i \tilde \mu_0)}{D_{-\nu-1}(\tilde \mu_0)}
\psi_\nu^{\hbox{\tiny out}}(\bm{r}) \hbox{\qquad (Dirichlet)}
\end{equation}
for Dirichlet boundary conditions and
\begin{equation}
\Phi(\bm{r}) = 
\psi_\nu^{\hbox{\tiny reg}}(\bm{r})
 - i^{\nu+1} \frac{D_{\nu}'(i \tilde \mu_0)}{D_{-\nu-1}'(\tilde \mu_0)}
\psi_\nu^{\hbox{\tiny out}}(\bm{r})
\hbox{\qquad (Neumann)}
\end{equation}
for Neumann boundary conditions, where prime denotes the derivative
of the parabolic cylinder function with respect to its argument and
$\nu=0,1,2,3,\cdots$.  These wavefunctions correspond to the scattering
$T$-matrix elements
${\cal T}_{\nu k_z \nu' k_z'} =
2\pi \delta(k_z - k_z')\delta_{\nu\nu'} {\cal T}_\nu^C$, with
\begin{eqnarray}
{\cal T}_\nu^C  &=&
-i^\nu  \frac{D_{\nu}(i\tilde \mu_0)}{D_{-\nu-1}(\tilde \mu_0)} 
\hbox{\qquad (Dirichlet),} \cr
{\cal T}_\nu^C &=& -i^{\nu+1} 
\frac{D_{\nu}'(i\tilde \mu_0)}{D_{-\nu-1}'(\tilde \mu_0)} \hbox{\qquad
(Neumann),}
\label{eqn:Tmat}
\end{eqnarray}
for the process where an incoming parabolic wave propagating in the
$+y$ direction is scattered into an outgoing parabolic wave
propagating radially.

The solutions we have obtained allowed us to construct the complete
free Green's function, the decomposition of a plane wave, and the
scattering $T$-matrices, which contain all the information we will
need to carry out our calculations.  However, we note that there also
exists a second set of solutions, representing the time-reversed
scattering process,
\begin{equation}
L(\lambda)M(\mu) = D_{-\nu-1}(i\tilde \lambda) D_{-\nu-1}(\tilde \mu) \,,
\label{eqn:regular2}
\end{equation}
for $\nu =0,1,2,3,\cdots$, which also are unchanged for $\lambda\to
-\lambda$ and $\mu\to -\mu$.  For real $k$, these solutions go
like $e^{-i\sqrt{k^2-k_z^2}y}$ and so propagate in the
$-y$ direction, in contrast to the solutions in
Eq.~(\ref{eqn:regular}), which go like $e^{i\sqrt{k^2-k_z^2}y}$
and propagate in the $+y$ direction.
We also have the the corresponding irregular solutions, 
\begin{equation}
L(\lambda)M(\mu) = D_{-\nu-1}(i\tilde \lambda) D_{\nu}(i\tilde \mu) \,,
\end{equation}
where again $\nu =0,1,2,3,\cdots$.  These solutions go
like $e^{-i\sqrt{k^2-k_z^2}r}$ and thus correspond to incoming radial
parabolic waves.

Since a decomposition of the Green's function like Eq.~(\ref{eqn:Green1})
typically consists of a sum over all scattering solutions, one might
wonder why this second set of solutions does not appear there.  One
can formally extend the sum to include all values of $\nu$, but for
$\nu <0$ the $\nu!$ in the denominator becomes a divergent gamma
function, so that these terms all gives zero contribution.  However,
we can also write the Green's function solely in terms of these
solutions as
\begin{eqnarray}
G(\lambda_1, \mu_1, z_1; \lambda_2, \mu_2, z_2, k) =
 \hspace*{0.7\linewidth}
\cr
\frac{1}{\sqrt{2\pi}}
\int_{-\infty}^\infty \frac{d k_z}{2 \pi} e^{i k_z(z_2 - z_1)}
\sum_{\nu=-\infty}^{-1} \frac{i^{-\nu-1}}{(-\nu-1)!}
D_{-\nu-1}(i\tilde \lambda_1)
D_{-\nu-1}(i\tilde \lambda_2)
D_{-\nu-1}(\tilde \mu_<)
D_{\nu}(i\tilde \mu_>) \,,
\label{eqn:Green2}
\end{eqnarray}
and the decomposition of the plane wave as
\begin{equation}
e^{i\bm{k}\cdot\bm{r}}
= e^{i k_z z} \frac{1}{\sin \frac{\phi}{2}}
\sum_{\nu=-\infty}^{-1} \frac{i^{-\nu-1}}{(-\nu-1)!}
\left(\cot \frac{\phi}{2}\right)^{(-\nu-1)}
D_{-\nu-1}(i\tilde \lambda) D_{-\nu-1}(\tilde \mu) \,,
\end{equation}
which now converges only for $k_y < 0$, since it consists only of
waves propagating downward.  Using these solutions, we could
construct the analogous scattering solutions for Neumann and Dirichlet
boundaries, which represent the process where an incoming radial
parabolic wave is scattered into an outgoing parabolic wave
propagating in the $-y$ direction.

\section{Parabolic Cylinder Opposite a Plane}\label{Pcyl-plane}

To calculate the Casimir force for a parabolic cylinder opposite a
plane, we will need an appropriate expression for the free Green's
function in terms of plane waves, and expansions translating between
these two bases.  For $y_2 > y_1$, the free Green's function can be
written in Cartesian coordinates as
\begin{equation}
G(\bm{r}_1,\bm{r}_2, k)=
\int_{-\infty}^\infty \frac{d k_z}{2 \pi} e^{i k_z(z_2 - z_1)}
\frac{i}{4\pi} \int_{-\infty}^\infty \frac{dk_x}{k_y} 
e^{i(k_x (x_2-x_1) + k_y (y_2 - y_1))} \,,
\label{eqn:Greenplane}
\end{equation}
where $k_y=\sqrt{k^2-k_x^2-k_z^2} = i\sqrt{\kappa^2+k_x^2+k_z^2}$.
We equate this expression to 
the Green's function in Eq.~(\ref{eqn:Green1}), expand the plane wave 
$\displaystyle e^{i \bm{k}\cdot{\bm{r}_2}}$ in Eq.~(\ref{eqn:Greenplane})
using Eq.~(\ref{eqn:plane1}), make the substitution $k_x \to -k_x$,
and then use the orthogonality of the
regular parabolic solutions to equate both sides term by term in the
sum over $\nu$.  The result is an expansion for the irregular
parabolic solutions in terms of plane waves:
\begin{equation}
\psi_\nu^{\hbox{\tiny out}}(\bm{r}) =
\int_{-\infty}^{\infty} dk_x \left[
\frac{i}{k_y \sqrt{8 \pi}}
\frac{\left(\tan \frac{\phi}{2}\right)^{\nu}}{\cos \frac{\phi}{2}}
\right]
e^{-i k_y y + i k_x x} e^{ik_z z} \,,
\label{eqn:expandout}
\end{equation}
which is valid for $y\leq 0$ and $\nu =0,1,2,3,\cdots$.  We have not
found this result in the previous literature,
though it is hinted at in \cite{Newman}.  The quantity in
brackets then defines the conversion matrix between
outgoing parabolic cylinder functions and plane waves propagating in
the $-y$ direction.  It allows us to propagate the outgoing waves
from the parabolic cylinder downward to the plane.  We
displace the origin of the Cartesian coordinates for the plane 
from the origin of the parabolic cylinder coordinates
by a distance $d$ in the $y$ direction, which simply introduces a
factor of $e^{ik_y d}$.

Due to invariance along the time and $z$ directions, we can make independent
computations for each $\kappa$ and $k_z$, and then integrate over both
quantities in the final result for the Casimir energy.  In the
scattering theory approach, the calculation can be formulated in terms
of scattering amplitudes by considering fluctuating multipoles
\cite{spheres,paraboloid}, or equivalently by using a generalized
$T$-operator formalism \cite{universal}.  In the latter approach,
which we adopt here, the ingredients we will need are the $T$-matrix
elements, the expansion of the outgoing wave in terms of plane waves,
and the normalization factors appearing in the Green's functions in
Eqs.~(\ref{eqn:Green1}) and (\ref{eqn:Greenplane}) \cite{universal}.  
The $T$-matrix elements for the parabolic cylinder are
given in Eq.~(\ref{eqn:Tmat}), and the $T$-matrix elements for
the plane are simply ${\cal T}^P_{k_x} = \pm 1$ for Neumann and
Dirichlet boundary conditions respectively.  Finally, we must include
the appropriate normalization factor \cite{universal}
$\displaystyle 
\frac{C^{\hbox{\tiny parabolic}}_\nu}{C^{\hbox{\tiny plane}}_{k_x}}$, 
where we can read off
$\displaystyle
C^{\hbox{\tiny parabolic}}_\nu = \sqrt{\frac{(-1)^\nu}{\nu! \sqrt{2 \pi}}}$
and $\displaystyle
C^{\hbox{\tiny plane}}_{k_x} = \sqrt{\frac{i}{4\pi k_y}}$
from the expressions for the free Green's function in 
Eqs.~(\ref{eqn:Green1}) and (\ref{eqn:Greenplane}).

We can then write the energy per unit length as
\begin{equation}
\frac{\cal E}{\hbar c L}=
\int_0^\infty \frac {d\kappa}{2 \pi} 
\int_{-\infty}^\infty \frac {dk_z}{2 \pi}
\log \det \left(\mathbbm{1}_{\nu \nu'} - 
{\cal T}_{\nu}^C
\int \frac{i d k_x}{2 k_y} (-1)^{(\nu - \nu')/2}
{\cal U}_{\nu k_x}(d) {\cal T}_{k_x}^P
\hat {\cal U}_{\nu ' k_x}(d)
\right)\,,
\end{equation}
where the matrix determinant runs over $\nu,\nu' =0,1,2,3,\cdots$.
Here we have defined the translation matrix
\begin{equation}
{\cal U}_{\nu k_x}(d) =
\frac{1}{\sqrt{\nu! \sqrt{2\pi}}}
\frac{\left(\tan \frac{\phi}{2}\right)^{\nu}}
{\cos \frac{\phi}{2}} e^{i k_y d} \,,
\label{eqn:cptrans}
\end{equation}
and, for convenience in later expressions in which we consider
different orientations of the parabolic cylinder, we have written
the reverse translation matrix as 
$\displaystyle {\cal U}_{\nu' k_x}(d)^\dagger = (-1)^{\nu'} 
\hat{\cal U}_{\nu' k_x}(d)$, with $\displaystyle 
\hat{\cal U}_{\nu k_x}(d) = {\cal U}_{\nu k_x}(d)$.
The complete energy per unit length is then
\begin{equation}
\frac{\cal E}{\hbar c L} = 
\int_0^\infty \frac {d\kappa}{2 \pi} 
\int_{-\infty}^\infty \frac {dk_z}{2 \pi}
\log \det \left(\mathbbm{1}_{\nu \nu'} - 
{\cal T}_{\nu}^C
\int_{-\infty}^\infty dk_x
{\cal T}^P_{k_x} \frac{i}{2\sqrt{2\pi}}
\frac{\left(\tan \frac{\phi}{2}\right)^{\nu+\nu'}}
{k_y \sqrt{\nu! \nu'!} \cos^2 \frac{\phi}{2}}
e^{2 i k_y d}\right) \,,
\label{eqn:cpenergy}
\end{equation}
where we have dropped a factor of $\displaystyle (-1)^{(\nu-\nu')/2}$
since it does not change the determinant.  We sum this result over
Dirichlet and Neumann boundary conditions to obtain the full
electromagnetic result.  This can be compared to the proximity force
approximation,
\begin{equation}
\frac{{\cal E}_{\hbox{\tiny pfa}}}{\hbar c L} = 
-\frac{\pi^2}{720}\int_{-\infty}^\infty dx 
\frac{1}{\left(d + \frac{1}{2}\left(\frac{x^2}{\mu_0} - \mu_0^2\right)
\right)^3} = -\frac{\pi^3}{240}\frac{\mu_0}{\left(2 d - \mu_0^2\right)^{5/2}
} \,,
\end{equation}
which is the sum of equal contributions from the Dirichlet and Neumann
cases.

We can make the following simplifications in Eq.~(\ref{eqn:cpenergy}):
\begin{itemize}

\item
The integral over $k_x$ is zero if $\nu + \nu'$ is odd, it
is symmetric in $\nu$, $\nu'$, and the integrand is even in $k_x$.

\item
We can replace $\displaystyle  \int_0^\infty \frac{d\kappa}{2\pi}
\int_{-\infty}^\infty \frac {dk_z}{2 \pi}$ by
$\displaystyle
\frac {1}{4 \pi} \int_0^\infty q dq$, where $q=\sqrt{\kappa^2 + k_z^2}$.

\item
We can further simplify the integral
\begin{equation}
I_{n,2d} = \int_{-\infty}^\infty dk_x \frac{i}{k_y} 
\frac{\left(\tan \frac{\phi}{2}\right)^{2n}}
{\cos^2 \frac{\phi}{2}} e^{2 i k_y d} \,,
\end{equation}
which appears in Eq.~(\ref{eqn:cpenergy}) with $\displaystyle
n=(\nu+\nu')/{2}$.  Here $n$ is always an integer, since the
translation matrix element vanishes if $\nu+\nu'$ is odd.  Setting
$\displaystyle u = \sqrt{1+\frac{k_x^2}{\kappa^2+k_z^2}}$, we have
\begin{equation}
I_{n,2d} = 4(-1)^{n} \int_1^\infty du \frac{(u-1)^{n-1/2}}{(u+1)^{n+3/2}}
e^{-2 u \sqrt{\kappa^2 + k_z^2} d} \,,
\end{equation}
which is given in terms of the confluent hypergeometric function of
the second kind $U(a,b,x)$ as
\begin{equation}
I_{n,2d} = 2(-1)^{n} e^{-2 \sqrt{\kappa^2 + k_z^2} d}
\Gamma\left(n+\frac{1}{2}\right)
U\left(n+\frac{1}{2}, 0, 4d \sqrt{\kappa^2 + k_z^2} \right)
= 2 \pi {\rm k}_{-2n-1}(2d\sqrt{\kappa^2 + k_z^2}) \,,
\label{eqn:Bateman}
\end{equation}
where ${\rm k}_\ell(x)$ is the Bateman ${\rm k}$-function
\cite{Bateman}.
\end{itemize}

\begin{figure}[htbp]
\includegraphics[width=0.5\linewidth]{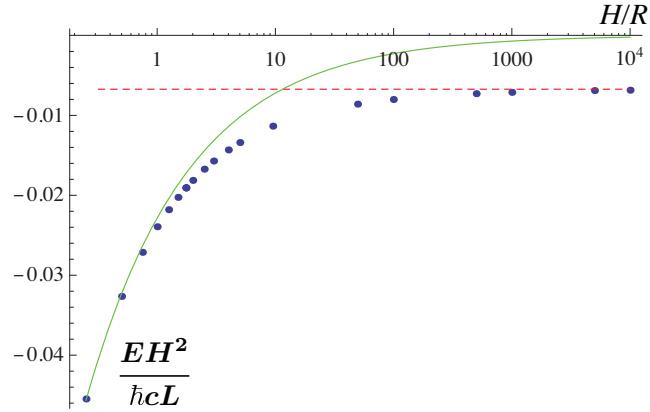}
\caption{The energy per unit length times $H^2$,
${\cal E} H^2/(\hbar c L)$, plotted versus $H/R$ for $\theta_C=0$ and
$R=1$ on a log-linear scale.  The dashed line gives the $R=0$ limit
and the solid curve gives the PFA result.}
\label{fig:EvsH}
\end{figure}

We now review some results that were reported previously using this
formalism \cite{paraboloid}.
To connect back to the physical configuration, it is convenient to
represent the final Casimir energy in terms of the radius of curvature
at the tip $R=\mu_0^2$ and the separation $H = d-R/2$.  
At small separations ($H/R\ll 1$) the proximity force approximation,
given by 
\begin{equation}
\frac{{\cal E}_{\hbox{\tiny pfa}}}{\hbar c L} = 
-\frac{\pi^2}{720} \int_{-\infty}^\infty
\frac{dx}{\left[H + x^2/(2R)\right]^3} = -\frac{\pi^3}{960\sqrt{2}}
\sqrt{\frac{R}{H^5}}\,,
\label{eq:pfa}
\end{equation}
should be valid.  The numerical results in Fig.~\ref{fig:EvsH} confirm
this expectation with a ratio of actual to PFA energy of $0.9961$ at
$H/R=0.25$ (with $R=1$).  We note that since the main contribution to
PFA is from the proximal parts of the two surfaces, the PFA result in
Eq.~(\ref{eq:pfa}) also applies to a circular cylinder with the same
radius $R$.  In the opposite limit, $R=0$, the parabolic cylinder
becomes a half-plane, and we can express the $T$-matrix in closed form
as well:
\begin{equation}
{\cal T}_{\nu}^C = 
-i^\nu \sqrt{\frac{2}{\pi}} \nu! \cos \frac{\nu \pi}{2}
\hbox{\quad (Dirichlet, $\mu_0=0$)},  \qquad \qquad
{\cal T}_{\nu}^C =
i^{\nu+1} \sqrt{\frac{2}{\pi}} \nu! \sin \frac{\nu \pi}{2}
\hbox{\quad (Neumann, $\mu_0=0$)} \,,
\end{equation}
the nonzero elements of which can be summarized compactly as
\begin{equation}
{\cal T}_{\nu}^C = -\sqrt{\frac{2}{\pi}} \nu! \,,
\end{equation}
where even $\nu$ corresponds to Dirichlet boundary conditions and odd
$\nu$ corresponds to Neumann boundary conditions.

We can thus write the full electromagnetic energy for the half-plane
perpendicular to the plane as
\begin{equation}
\frac{\cal E}{\hbar c L} = \frac{1}{4\pi} \int_0^\infty q dq \log \det
\left(\mathbbm{1}_{\nu \nu'} -   (-1)^\nu {\rm k}_{-\nu-\nu'-1}(2 q H)
\right) = -\frac{C_\perp}{H^2},
\label{eq:zeroR}
\end{equation}
where the Bateman ${\rm k}$-function is nonzero only for $\nu + \nu'$
even, and we have dropped factors that cancel in the determinant.
The factor of $(-1)^\nu$ in this expression arises from the $T$-matrix
element for the plane.  Numerically, we find $C_\perp=0.0067415$,
which is shown by a dashed line in Fig.~\ref{fig:EvsH}.

This geometry was studied using the world-line method for a scalar field
with {\em Dirichlet} boundary conditions in Ref.~\cite{Gies}.  (The
world-line approach requires a large-scale numerical computation,
and it is not known how to extend this method beyond the case of a
scalar with Dirichlet boundary conditions).  In our calculation, the
Dirichlet component of the electromagnetic field makes a contribution
$C^D_\perp=0.0060485$ to our result, in reasonable agreement with the
value of  $C^D_\perp=0.00600(2)$ in Ref.~\cite{Gies}.  These results
are also in agreement with the calculation in Ref.~\cite{Kabat}.

\begin{figure}[htbp]
\includegraphics[width=0.4\linewidth]{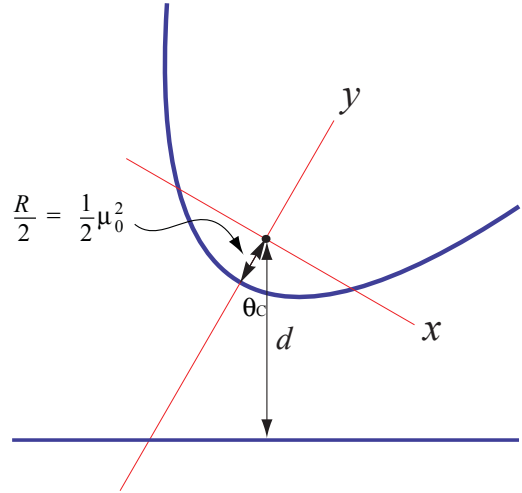}
\caption{The geometry of a tilted parabolic cylinder in front of a plane.}
\label{fig:tilt}
\end{figure}

It is straightforward to extend this calculation to the case where the
parabolic cylinder (of any radius) is rotated by an angle $\theta_C$
around its focal axis, as shown in Fig.~\ref{fig:tilt}.  In place of
Eq.~(\ref{eqn:cptrans}), we have
\begin{equation}
{\cal U}_{\nu k_x}(d, \theta_C) = 
\frac{1}{\sqrt{\nu! \sqrt{2\pi}}}
\frac{\left(\tan \frac{\phi + \theta_C}{2}\right)^{\nu}}
{\cos \frac{\phi + \theta_C}{2}} e^{i k_y d}
\hbox{\quad and \quad}
\hat{\cal U}_{\nu k_x}(d, \theta_C) =
{\cal U}_{\nu k_x}(d, -\theta_C) \,.  
\label{eqn:cptransrot}
\end{equation}
However, now the integral over $k_x$ is not symmetric, and the
matrix elements with $\nu+\nu'$ odd need not vanish.
\begin{figure}[htbp]
\includegraphics[width=0.5\linewidth]{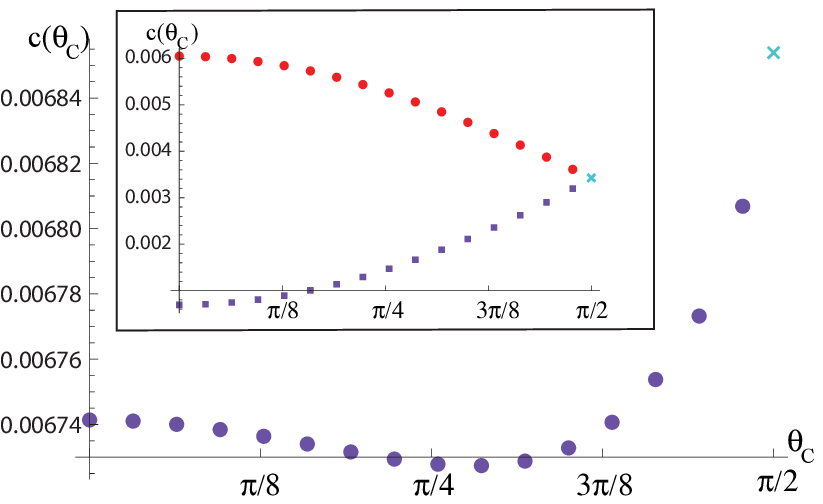}
\includegraphics[width=0.45\linewidth]{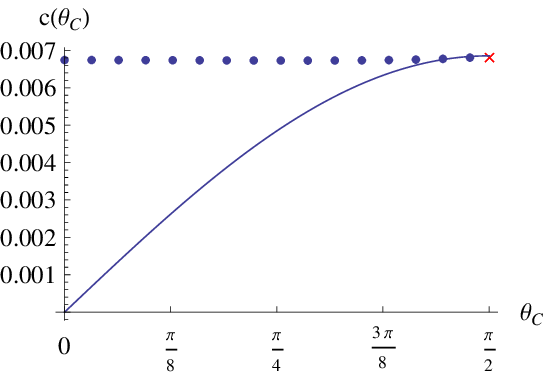}
\caption{The dependence of the Casimir energy on the tilt angle for a
half-plane opposite a plane.  The half-plane is a parabolic cylinder
with $R=0$, which is oriented perpendicular to the plane for
$\theta_C=0$ and parallel to the plane for $\theta_C=\pi/2$.
The left panel shows the coefficient $c(\theta_C)$ (see text) as a function of
$\theta_C$, with the exact parallel plate result at $\theta_C=\pi/2$
marked with a cross.  The inset shows the Dirichlet (circles) and
Neumann (squares) contributions to the full electromagnetic result.
The right panel again shows $c(\theta_C)$, but now in comparison to
the proximity force approximation (solid line).  Note the large
discrepancy between the PFA and the exact results as $\theta_C\to 0$.
}
\label{fig:cAmp}
\end{figure}

We again consider the $R\to0$ limit in analyzing this result.
From dimensional analysis, the electromagnetic Casimir energy at
$R=0$ takes the now $\theta_C$-dependent form 
\begin{equation}
\frac{{\cal E}}{\hbar c L} = -\frac{C(\theta_C)}{H^2} \,,
\label{eq:Etheta}
\end{equation}
where $H=d$ for $R=0$.  Following Ref.~\cite{Gies},
which considers the Casimir energy for a scalar
field with Dirichlet boundary conditions in this geometry,
we plot $c(\theta_C)=\cos(\theta_C)C(\theta_C)$ in Fig.~\ref{fig:cAmp}.
A particularly interesting limit is $\theta_C\to \pi/2$,
as the two plates become parallel.  In this case, the leading
contribution to the Casimir energy should be proportional to the area
of the half-plane according to the parallel plate formula,
${\cal E}_\parallel /(\hbar cA)= -c_\parallel/H^3$ with
$c_\parallel=\pi^2/720$, plus a subleading correction due to the
edge.  Multiplying by $\cos\theta_C$ has removed the divergence in 
$C(\theta_C)$ as $\theta_C\to \pi/2$.  As in Ref.~\cite{Gies}, we assume 
$c(\theta_C\to \pi/2)=c_\parallel/2+ \left(\theta_C - \pi/2\right)
c_{\hbox{\tiny edge}}$, although we cannot rule out the possibility of
additional non-analytic forms, such as logarithmic or other
singularities.  With this assumption, we can estimate
the edge correction $c_{\hbox{\tiny edge}} = 0.0009$ from the data in
Fig.~\ref{fig:cAmp}.  From the inset in Fig.~\ref{fig:cAmp}, we
estimate the Dirichlet and Neumann contributions to this result to be
$c_{\hbox{\tiny edge}}^D = -0.0025$ (in agreement with
\cite{Gies} within our error estimates) and 
$c_{\hbox{\tiny edge}}^N = 0.0034$ respectively.  Because higher partial
waves become more important as $\theta_C \to \pi/2$, reflecting the
divergence in $C(\theta_C)$ in this limit, we have used larger values of
$\nu_{\hbox{\tiny max}}$ for $\theta_C$ near $\pi/2$.  In
Fig.~\ref{fig:cAmp} we also show a comparison to the proximity force
approximation.  The PFA is clearly of no use at $\theta_C=0$, since it
simply gives zero, while at $\theta_C=\pi/2$ the PFA gives the correct
energy but incorrectly has zero slope, since it misses the edge correction.

We have found that the edge correction is small in the
electromagnetic case, as a result of the near-cancellation between the
Dirichlet and Neumann contributions.  By using Babinet's
principle, it is possible to show that this suppression of edge
effects is a general feature of any thin conductor, arising because the
leading term in the multiple reflection expansion is identically zero
\cite{Babinet}.

\section{Two Parabolic Cylinders}\label{2Pcyls}

\begin{figure}
\hfil
\includegraphics[width=0.25\linewidth]{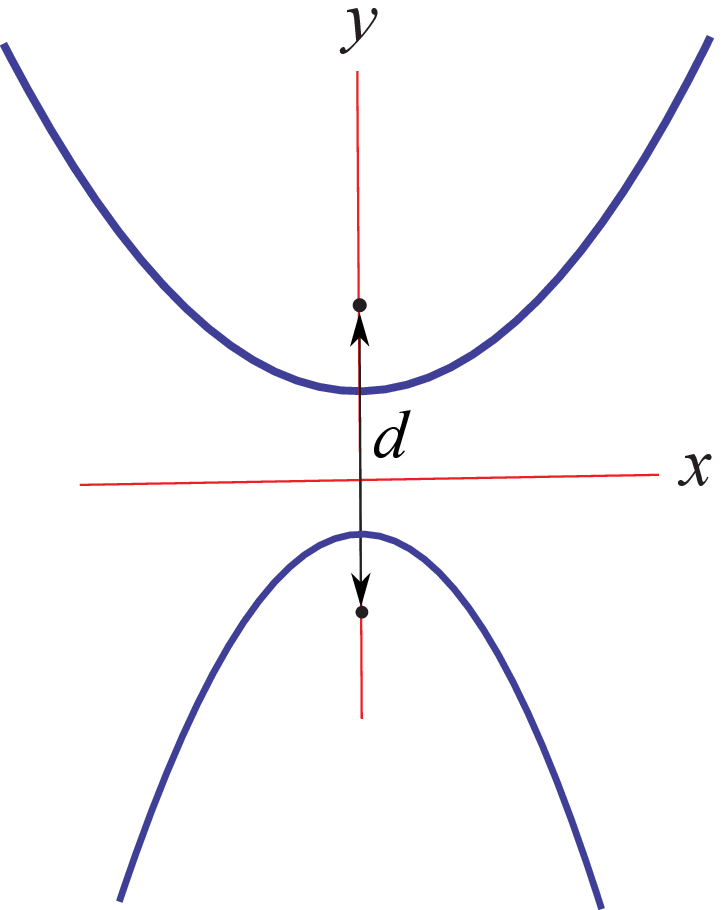}
\hfil
\includegraphics[width=0.3\linewidth]{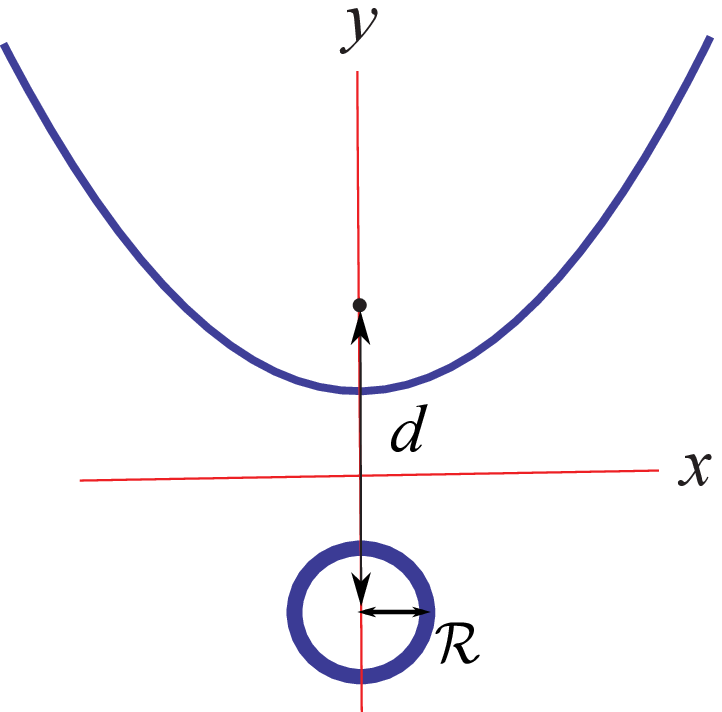}
\hfil
\caption{Exterior parabolic cylinder geometries:  Two parabolic cylinders
outside one another (left panel) and an ordinary cylinder outside a
parabolic cylinder (right panel).}
\label{fig:parageomout}
\end{figure}

We next consider the force between two parabolic cylinders 
opening in opposite directions, as shown in the left panel of
Fig.~\ref{fig:parageomout}.  We will consider the generalization to
arbitrary orientation below.  We need to express the outgoing waves
from one parabolic cylinder in terms of the regular waves for the other. 
We let $\bar{\bm{r}}$ represent the coordinates of the second
parabolic cylinder, $x = \bar x$, $y = -\bar y - d$, and $z= \bar z$.
Using Eq.~(\ref{eqn:expandout}) for $\bar{\bm{r}}$, we have
\begin{equation}
\psi_\nu^{\hbox{\tiny out}}(\bar{\bm{r}})
= e^{ik_z z} \frac{1}{\sqrt{8\pi}} \int_{-\infty}^{\infty} dk_x
\frac{i}{k_y}
\frac{\left(\tan \frac{\phi}{2}\right)^{\nu}}{\cos \frac{\phi}{2}}
e^{i k_x x + i k_y y + i k_y d} \,.
\label{eq:outreg}
\end{equation}

Now we use the expansion of the plane wave, Eq.~(\ref{eqn:plane1}), to
obtain
\begin{eqnarray}
\psi_\nu^{\hbox{\tiny out}}(\bar{\bm{r}}) 
= \sum_{\nu'=0}^\infty  \left[
\frac{1}{\nu'!\sqrt{8\pi}} \int_{-\infty}^{\infty} dk_x
\frac{i}{k_y}
\frac{\left(\tan \frac{\phi}{2}\right)^{\nu+\nu'}}{\cos^2 \frac{\phi}{2}}
e^{i k_y d} \right]
\psi_{\nu'}^{\hbox{\tiny reg}}(\bm{r}) \,,
\end{eqnarray}
where $d$ is the interfocal separation.  We can then obtain the
translation coefficient from the quantity in brackets.  We thus obtain
the Casimir interaction energy
\begin{equation}
\frac{\cal E}{\hbar c L} = 
\int_0^\infty \frac {d\kappa}{2 \pi} 
\int_{-\infty}^\infty \frac {dk_z}{2 \pi}
\log \det \left(\mathbbm{1}_{\nu \nu'} - 
{\cal T}_{\nu}^C \sum_{\nu''=0}^\infty
{\cal U}_{\nu \nu''}(d)
{\cal T}_{\nu''}^{\bar C}
\hat{\cal U}_{\nu'' \nu'}(d)
\right)  \,,
\end{equation}
where ${\cal T}^{C}$ and ${\cal T}^{\bar C}$ are the scattering
$T$-matrix elements for the two parabolic cylinders (which can have
different radii), the translation matrix elements are given by
\begin{equation}
{\cal U}_{\nu \nu'}(d) = \hat {\cal U}_{\nu \nu'}(d) = 
\frac{1}{\sqrt{8 \pi \nu! \nu'!}}
\int_{-\infty}^\infty dk_x \frac{i}{k_y} 
\frac{\left(\tan \frac{\phi}{2}\right)^{\nu+\nu'}}
{\cos^2 \frac{\phi}{2}} e^{i k_y d} \,,
\label{eqn:cctrans}
\end{equation}
and the determinant runs over $\nu,\nu'=0,1,2,3,\cdots$.  
Here again we have defined $\displaystyle \hat {\cal U}_{\nu \nu'}(d)
= (-1)^{\nu + \nu'} {\cal U}_{\nu \nu'}(d)^\dagger$, where
$\displaystyle {\cal U}_{\nu \nu'}(d)^\dagger$ is the reverse
translation matrix.  We sum the results for Dirichlet and
Neumann boundary conditions to obtain the result for electromagnetism.
The analogous numerical simplifications apply here as in the case of
the plane, and we can use Eq.~(\ref{eqn:Bateman}), now with $d$
instead of $2d$, to express the translation matrix elements in
Eq.~(\ref{eqn:cctrans}) in terms of the Bateman ${\rm k}$-function.
The extension to the tilted case is also analogous; now the
angle of rotation can be different for the two translation matrices,
corresponding to different angles of rotation for the two parabolic
cylinders.  We can also introduce a translation in the $x$-direction
$d_x$, in addition to the existing translation $d$ in the
$y$-direction.  For rotations $\theta_C$ and $\bar\theta_C$ of the two
parabolic cylinders and $x$-translation $d_x$, we have 
\begin{eqnarray}
{\cal U}_{\nu \nu'}(d,\theta_C, \bar\theta_C, d_x) &=& 
\frac{1}{\sqrt{8 \pi \nu! \nu'!}}
\int_{-\infty}^\infty dk_x \frac{i}{k_y} 
\frac{\left(\tan \frac{\phi + \theta_C}{2}\right)^{\nu}}
{\cos \frac{\phi + \theta_C}{2}}
\frac{\left(\tan \frac{\phi + \bar\theta_C}{2}\right)^{\nu'}}
{\cos \frac{\phi + \bar\theta_C}{2}} e^{i k_y d} e^{i k_x d_x} \,, \cr
\hat {\cal U}_{\nu \nu'}(d,\theta_C, \bar\theta_C, d_x) &=&
{\cal U}_{\nu \nu'}(d,-\theta_C, -\bar\theta_C, -d_x) \,,
\label{eqn:trans2par}
\end{eqnarray}
where we must have $d>0$, but $d_x$ can have either sign, representing
a translation in either horizontal direction.

\begin{figure}[htbp]
\includegraphics[width=0.3\linewidth]{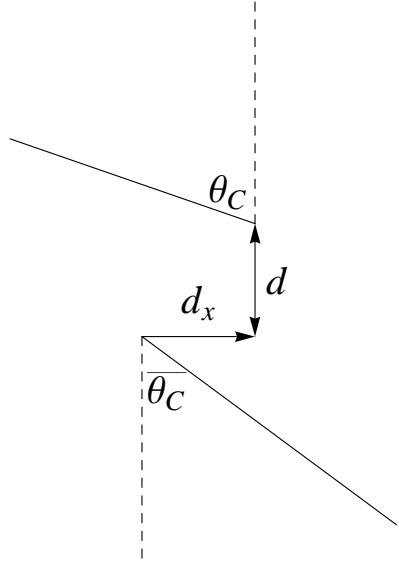}
\caption{Two half--planes tilted by angles $\theta_C$ and
$\overline{\theta}_C$, and displaced by $d_x$ and $d$.}
\label{fig:planegeom}
\end{figure}

\begin{figure}[htbp]
\includegraphics[width=0.5\linewidth]{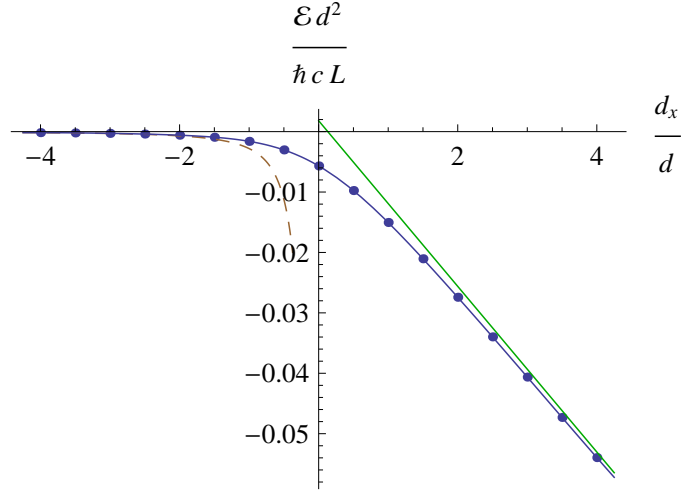}
\caption{Electromagnetic Casimir interaction energy per unit length
for overlapping planes as a function of horizontal displacement, in
units of the vertical separation $d$.  Solid points are obtained from the
exact calculation described in the text.  The solid line connecting
them is a rational function fit to guide the eye.  The dashed line
gives the energy for the limit where the planes are edge-to-edge,
while the solid straight line gives the standard parallel plate result
for the overlap area, plus edge corrections.}
\label{fig:overlap}
\end{figure}

By considering two parabolic cylinders of zero radius, we can study
the Casimir interactions of two half-planes, as illustrated in
Fig.~\ref{fig:planegeom}.  These techniques, together with a
multiple reflection expansion, were used in Ref.~\cite{planes} to
obtain a variety of results in half-plane geometries.
We take $\theta_C=\bar\theta_C=\pi/2$, so that we are considering
parallel half-planes, where positive $d_x$ gives the width of the region
over which they overlap, while negative $d_x$ gives a horizontal
displacement of the edges away from each other.  In this case,
Eq.~(\ref{eqn:trans2par}) simplifies to
\begin{equation}
{\cal U}_{\nu \nu'}\left(d, \frac{\pi}{2}, \frac{\pi}{2}, d_x\right)
= \hat {\cal U}_{\nu \nu'}\left(d, \frac{\pi}{2}, \frac{\pi}{2}, d_x\right)
= \frac{1}{\sqrt{2 \pi \nu! \nu'!}}
\int_{-\infty}^\infty dv 
\frac{(1-iv)^{(\nu+\nu'-1)/2}}{(1+iv)^{(\nu+\nu'+3)/2}}
e^{i d_x v \sqrt{\kappa^2 + k_z^2}}
e^{-d \sqrt{v^2+1} \sqrt{\kappa^2 + k_z^2}}\,.
\end{equation}
Results are shown in Fig.~\ref{fig:overlap}, along with
approximations valid in two limiting cases:  First, for $d_x$ very
negative, we can ignore the vertical displacement.  The configuration
is then equivalent to the case of $\theta_C=\bar\theta_C=0$, which
gives ${\cal E}/(\hbar c L) = -0.0020856/d^2$ at separation $d$, which
is shown as a dashed line in Fig.~\ref{fig:overlap}.  Second, for
$d_x$ large and positive, we can take the standard result for parallel
plates  ${\cal E}_\parallel/(\hbar c L) = -\pi^2 d_x/(720 d^3)$ plus
twice the edge correction ${\cal E}_{\hbox{\tiny edge}}/(\hbar c L) =
0.0009/d^2$ found above for a half-plane parallel to a plane, which is
shown as a solid line in Fig.~\ref{fig:overlap}.

As these examples illustrate, our description of the two half-planes
is redundant:  Different parameter choices lead to the same physical 
configuration, a property we have used to check our calculations.
The numerical convergence of physically equivalent configurations 
can be quite different, however.  For example, in the case of
$\theta=\bar\theta = 0$, when both $d_x$ and $d$ increase, the
Casimir interaction energy decreases, since the half-planes are becoming
further apart.  In the scattering bases we have chosen, however, in
$d$ this effect appears directly through a decaying exponential, while in
$d_x$ it appears through the cancellation of an oscillating integrand.
As a result, we need to maintain $d>0$, but can consider either sign
of $d_x$.

\section{Parabolic Cylinder and Ordinary cylinder}\label{Pcyl-cyl}

We next consider the case of an ordinary cylinder outside a
parabolic cylinder, as shown in the right panel of
Fig.~\ref{fig:parageomout}.  In ordinary cylindrical coordinates, we
have regular solutions given in terms of Bessel functions, $e^{i k_z
z} e^{i\ell \theta} J_\ell(\sqrt{k^2-k_z^2} r)$, and outgoing
solutions given in terms of Hankel functions of the first kind, $e^{i
k_z z} e^{i\ell \theta} H^{(1)}_\ell(\sqrt{k^2-k_z^2} r)$, both
indexed by angular momentum $\ell$.  We use the expansion of a plane
wave in regular ordinary cylindrical wavefunctions,
\begin{equation}
e^{i\bm{k} \cdot \bm{r}} = 
e^{i k_z z} \sum_{\ell=-\infty}^{\infty}
e^{i\ell \phi} e^{i\ell \theta} J_\ell(\sqrt{k^2-k_z^2} r) \,,
\label{eq:cylplane}
\end{equation}
where $\phi$ is defined as in Eq.~(\ref{eq:phi}) and $\theta$ and $r$ are the
ordinary cylindrical coordinates for $\bm{r}$.  In these coordinates,
the free Green's function is given by
\begin{equation}
G(\bm{r}_1, \bm{r}_2, k)
= \frac{i}{4}\int_{-\infty}^\infty \frac{d k_z}{2 \pi}
e^{ik_z(z_1 - z_2)} \sum_{\ell=-\infty}^\infty 
e^{i\ell (\theta_1-\theta_2)} J_\ell(\sqrt{k^2-k_z^2} r_<)
H^{(1)}_\ell(\sqrt{k^2-k_z^2} r_>) \,,
\label{eqn:Greencyl}
\end{equation}
where $r_<$ ($r_>$) is the smaller (larger) of $r_1$ and $r_2$.

The $T$-matrix elements for an ordinary cylinder of radius ${\cal R}$
are given in terms of Bessel and Hankel functions and their modified
counterparts by
$\displaystyle {\cal T}_{\ell k_z \ell' k_z'} = 
2\pi \delta(k_z - k_z')\delta_{\ell\ell'} {\cal T}_\ell^O$, with
\begin{eqnarray}
{\cal T}_\ell^O  &=&
-\frac{J_\ell \left( {\cal R} \sqrt{k^2 - k_z^2}\right)}
{H_\ell^{(1)}\left({\cal R} \sqrt{k^2 - k_z^2}\right)}
= -\frac{\pi}{2} i^{2\ell+1}
\frac{I_\ell \left( {\cal R} \sqrt{\kappa^2 + k_z^2}\right)}
{K_\ell\left({\cal R} \sqrt{\kappa^2 + k_z^2}\right)}
\hbox{\qquad (Dirichlet),} \cr
{\cal T}_\ell^O  &=&
-\frac{J_\ell' \left( {\cal R} \sqrt{k^2 - k_z^2}\right)}
{H_\ell^{(1)}{}'\left({\cal R} \sqrt{k^2 - k_z^2}\right)}
= -\frac{\pi}{2} i^{2\ell+1}
\frac{I_\ell' \left( {\cal R} \sqrt{\kappa^2 + k_z^2}\right)}
{K_\ell'\left({\cal R} \sqrt{\kappa^2 + k_z^2}\right)}
\hbox{\quad (Neumann),}
\end{eqnarray}
where prime indicates a derivative with respect to the function's
argument.

For an ordinary cylinder outside a parabolic cylinder with separation
$d$ between the center of the ordinary cylinder and the focus of the
parabolic cylinder, we substitute Eq.~(\ref{eq:cylplane}) into
Eq.~(\ref{eq:outreg}) to obtain
\begin{eqnarray}
\psi_\nu^{\hbox{\tiny out}}(\bar{\bm{r}})
&=& e^{ik_z z} \frac{1}{\sqrt{8\pi}} \int_{-\infty}^{\infty} dk_x
\frac{i}{k_y}
\frac{\left(\tan \frac{\phi}{2}\right)^{\nu}}{\cos \frac{\phi}{2}}
e^{i k_y d} \sum_{\ell=-\infty}^{\infty}
e^{i\ell \phi} e^{i\ell \theta} J_\ell(\sqrt{k^2-k_z^2} r) \cr
&=& \sum_{\ell=-\infty}^{\infty} 
\left[\frac{1}{\sqrt{8\pi}} 
\int_{-\infty}^{\infty} dk_x \frac{i}{k_y} e^{i\ell \phi} 
\frac{\left(\tan \frac{\phi}{2}\right)^{\nu}}{\cos \frac{\phi}{2}}
e^{i k_y d} \right]
e^{ik_z z} e^{i\ell \theta} J_\ell(\sqrt{k^2-k_z^2} r) \,,
\end{eqnarray}
where again the quantity in brackets is the coefficient we need to
compute the translation matrix.

We thus obtain the Casimir interaction energy
\begin{equation}
\frac{\cal E}{\hbar c L} = 
\int_0^\infty \frac {d\kappa}{2 \pi} 
\int_{-\infty}^\infty \frac {dk_z}{2 \pi}
\log \det \left(\mathbbm{1}_{\nu \nu'} - 
{\cal T}_{\nu}^C
\sum_{\ell=-\infty}^\infty
{\cal U}_{\nu \ell}(d)
{\cal T}_{\ell}^{O}
\hat{\cal U}_{\nu' \ell}(d)
\right)  \,,
\end{equation}
where
\begin{equation}
{\cal U}_{\nu \ell}(d) = \hat {\cal U}_{\nu \ell}(d) = 
\frac{1}{\sqrt{\nu!\sqrt{2\pi}}} \sqrt{\frac{4(-1)^\ell}{i}}
\frac{1}{\sqrt{8\pi}} 
\int_{-\infty}^{\infty} dk_x \frac{i}{k_y} e^{i\ell \phi} 
\frac{\left(\tan \frac{\phi}{2}\right)^{\nu}}{\cos \frac{\phi}{2}}
e^{i k_y d} 
\end{equation}
and the determinant again runs over $\nu,\nu'=0,1,2,3,\cdots$.
As before, for convenience we have defined
$\displaystyle \hat {\cal U}_{\nu \ell}(d) = (-1)^{\nu + \ell}
{\cal U}_{\nu \ell}(d)^\dagger$, where $\displaystyle {\cal U}_{\nu
\ell}(d)^\dagger$ is the reverse translation matrix.
We can again simplify this expression for numerical computation by
combining the $\kappa$ and $k_z$ integrals, and by
exploiting symmetries in $\ell\to -\ell$ and $k_x \to -k_x$.
Generalizations to include horizontal translation and tilt of the
parabolic cylinder also work in the same way as before,
and taking the limit in which the radius of the parabolic cylinder goes
to zero gives the Casimir energy for a cylindrical wire opposite a
half-plane. 

\section{Interior Geometries}\label{interior}

Up to now, we have considered ``exterior'' geometries in which the
objects are outside one another.  However, simple modifications of
these techniques enable us to also consider ``interior'' geometries
using the formalism of Ref.~\cite{interior}.  (Interior geometries
were also considered in \cite{Lombardo}, using large-scale
computation.)  There are two changes required for this case: The
$T$-matrix of the outside object must be inverted, and we require the
translation matrix connecting the regular solutions for the different
objects, rather than the one connecting outgoing solutions for one
object to regular solutions for the other.

\begin{figure}
\hfil
\includegraphics[width=0.25\linewidth]{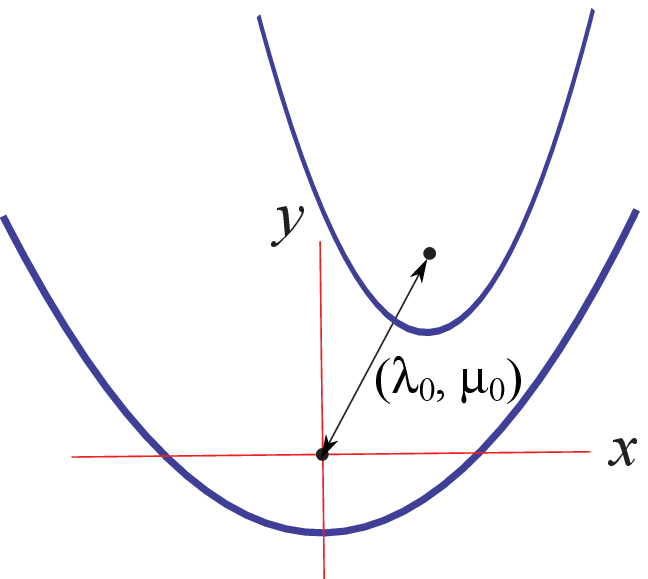}
\hfil
\includegraphics[width=0.3\linewidth]{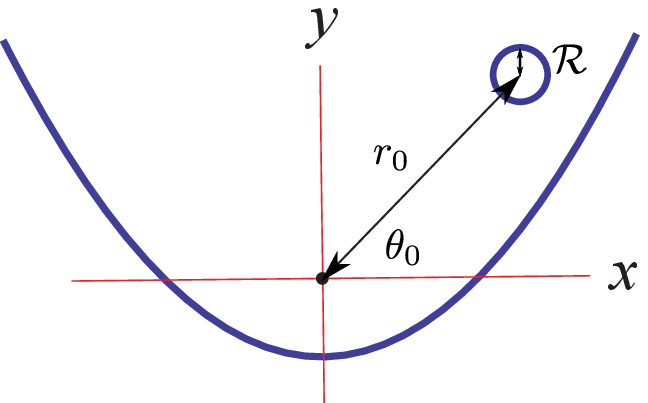}
\hfil
\caption{Interior parabolic cylinder geometries:  Two parabolic cylinders
inside one another (left panel) and an ordinary cylinder inside a
parabolic cylinder (right panel).}
\label{fig:parageom}
\end{figure}

We first consider two parabolic cylinders inside one another.  We
parameterize the displacement between their foci in parabolic
cylinder coordinates by $\lambda_0$ and $\mu_0$, as shown
in the left panel of Fig.~\ref{fig:parageom}.
Following Ref.~\cite{Epstein}, we can derive the translation matrix
for regular solutions appropriate to the inside problem.  Let 
$\bm{r}' = \bm{r} + \bm{r}_0$ and consider the equation 
$e^{i\bm{k}\cdot \bm{r}'} = e^{i\bm{k}\cdot \bm{r}_0}
e^{i\bm{k}\cdot \bm{r}}$.  Using Eq.~(\ref{eqn:plane1}), we have
\begin{eqnarray}
\frac{1}{\cos \frac{\phi}{2}}
\sum_{\nu'=0}^\infty \frac{1}{\nu'!}
\left(\tan \frac{\phi}{2} \right)^{\nu'}
\psi_{\nu'}^{\hbox{\tiny reg}} (\lambda', \mu') 
&=& \frac{1}{\cos \frac{\phi}{2}}
\sum_{\nu_0=0}^\infty \frac{1}{\nu_0!}
\left(\tan \frac{\phi}{2} \right)^{\nu_0}
\psi_{\nu_0}^{\hbox{\tiny reg}} (\lambda_0, \mu_0) \cr
&\times& \frac{1}{\cos \frac{\phi}{2}}
\sum_{\nu=0}^\infty \frac{1}{\nu!}
\left(\tan \frac{\phi}{2} \right)^{\nu}
\psi_{\nu}^{\hbox{\tiny reg}} (\lambda, \mu) \,.
\end{eqnarray}
Now we let $\displaystyle t=\tan \frac{\phi}{2}$, so 
$\displaystyle \cos \frac{\phi}{2} = \frac{1}{\sqrt{1+t^2}}$, and
consider the case where $|t|<1$ to obtain
\begin{equation}
\sum_{\nu'=0}^\infty \frac{t^{\nu'}}{\nu'!}
\psi_{\nu'}^{\hbox{\tiny reg}} (\lambda', \mu') 
= \sqrt{1+t^2} \sum_{\nu_0=0}^\infty \frac{t^{\nu_0}}{\nu_0!}
\psi_{\nu_0}^{\hbox{\tiny reg}} (\lambda_0, \mu_0)
\sum_{\nu=0}^\infty \frac{t^\nu}{\nu!}
\psi_{\nu}^{\hbox{\tiny reg}} (\lambda, \mu) \,.
\label{eqn:addition}
\end{equation}
Next we write
\begin{equation}
\sqrt{1+t^2} = \sum_{n=0}^\infty \alpha_n t^n\,,
\hbox{\quad where \quad}
\alpha_n = \left\{ \begin{array}{l@{\quad}l}
0 & \hbox{if $n$ is odd} \cr
\frac{(-1)^{n/2} \Gamma\left(\frac{n-1}{2}\right)}
{\Gamma\left(-\frac{1}{2}\right)\frac{n}{2}!} &
\hbox{if $n$ is even}
\end{array} \right. \,.
\end{equation}
Substituting this result into Eq.~(\ref{eqn:addition}) and equating
powers of $t$ results in
\begin{equation}
\psi_{\nu'}^{\hbox{\tiny reg}} (\lambda', \mu') 
= \sum_{\nu=0}^{\nu'} \left[\sum_{\nu_0=0}^{\nu'-\nu}
\frac{\nu'!}{\nu! \nu_0!} \alpha_{\nu'-\nu-\nu_0}
\psi_{\nu_0}^{\hbox{\tiny reg}} (\lambda_0, \mu_0)\right]
\psi_{\nu}^{\hbox{\tiny reg}} (\lambda, \mu)  \,,
\end{equation}
which yields the coefficient we need from the quantity in brackets.

The Casimir interaction energy is then
\begin{equation}
\frac{\cal E}{\hbar c L} = 
\int_0^\infty \frac {d\kappa}{2 \pi} 
\int_{-\infty}^\infty \frac {dk_z}{2 \pi}
\log \det \left(\mathbbm{1}_{\nu \nu'} - 
\left({\cal T}_{\nu}^C\right)^{-1}
{\cal V}_{\nu \nu'}(\lambda_0, \mu_0)
{\cal T}_{\nu'}^{\bar C}
\hat{\cal V}_{\nu \nu'}(\lambda_0, \mu_0)
\right)  \,,
\end{equation}
where ${\cal T}_{\nu}^{C}$ (${\cal T}_{\nu}^{\bar C}$) is the $T$-matrix
for the outer (inner) parabolic cylinder,
\begin{equation}
{\cal V}_{\nu \nu'}(\lambda_0, \mu_0) = 
\hat{\cal V}_{\nu \nu'}(-\lambda_0, \mu_0) = 
\sum_{\nu_0=0}^{\nu'-\nu}
\frac{1}{\nu_0!} \alpha_{\nu'-\nu-\nu_0}
\psi_{\nu_0}^{\hbox{\tiny reg}} (\lambda_0, \mu_0) \,,
\end{equation}
and we have dropped normalization factors that cancel in the
determinant.

For an ordinary cylinder inside a parabolic cylinder, 
as shown in the right panel of Fig.~\ref{fig:parageom}, we
again let $\bm{r}' = \bm{r} + \bm{r}_0$ and consider the equation 
$e^{i\bm{k}\cdot \bm{r}'} = e^{i\bm{k}\cdot \bm{r}_0}
e^{i\bm{k}\cdot \bm{r}}$, but now we expand the left-hand side in
parabolic cylinder coordinates and the right-hand side in ordinary
cylindrical coordinates, to obtain
\begin{equation}
\frac{1}{\cos \frac{\phi}{2}}\sum_{\nu'=0}^\infty 
\frac{\left(\tan \frac{\phi}{2} \right)^{\nu'}}{\nu'!}
\psi_{\nu'}^{\hbox{\tiny reg}} (\lambda', \mu')
= e^{i k_z z_0} \sum_{\ell_0=-\infty}^{\infty}
e^{i\ell_0 \phi} e^{i\ell_0 \theta_0} J_{\ell_0}(\sqrt{k^2-k_z^2} r_0)
e^{i k_z z} \sum_{\ell=-\infty}^{\infty}
e^{i\ell \phi} e^{i\ell \theta} J_{\ell}(\sqrt{k^2-k_z^2} r) \,.
\end{equation}
As before, setting $\displaystyle t = \tan \frac{\phi}{2}$, so that
$\displaystyle \sqrt{1+t^2} = \frac{1}{\cos \frac{\phi}{2}}$
and $\displaystyle \phi = \frac{1}{i} \log \frac{1+it}{1-it}$, yields
\begin{equation}
\sum_{\nu'=0}^\infty  \frac{t^{\nu'}}{\nu'!}
\psi_{\nu'}^{\hbox{\tiny reg}} (\lambda', \mu')
= \frac{1}{\sqrt{1+t^2}}
\sum_{\ell=-\infty}^{\infty}
\sum_{\ell_0=-\infty}^{\infty}
\left(\frac{1+it}{1-it}\right)^{\ell+\ell_0}
e^{i k_z z_0} e^{i\ell_0 \theta_0} J_{\ell_0}(\sqrt{k^2-k_z^2} r_0)
e^{i k_z z} e^{i\ell \theta} J_{\ell}(\sqrt{k^2-k_z^2} r) \,.
\end{equation}
We take $\nu$ derivatives with respect to $t$ and then set
$t=0$ to obtain
\begin{equation}
\psi_{\nu}^{\hbox{\tiny reg}} (\lambda', \mu') = 
\sum_{\ell=-\infty}^{\infty} \left[
\sum_{\ell_0=-\infty}^{\infty}
\frac{d^{\nu}}{dt^{\nu}}\left( \frac{1}{\sqrt{1+t^2}}
\left(\frac{1+it}{1-it}\right)^{\ell+\ell_0} \right|_{t=0}
e^{i k_z z_0} e^{i\ell_0 \theta_0} J_{\ell_0}(\sqrt{k^2-k_z^2} r_0) \right]
e^{i k_z z} e^{i\ell \theta} J_{\ell}(\sqrt{k^2-k_z^2} r) \,,
\end{equation}
where again the quantity in brackets will give the coefficient we
need.  Using the generalized binomial expansion, we obtain
\begin{equation}
\beta_{\nu, \ell} =
\frac{d^{\nu}}{dt^{\nu}}\left( \frac{1}{\sqrt{1+t^2}}
\left(\frac{1+it}{1-it}\right)^{\ell} \right|_{t=0}
= \nu! \sum_{\genfrac{}{}{0pt}{1}{n=0}
{n\equiv\,\nu \bmod 2}}^{\min(2|\ell|, \nu)}
\frac{(2|\ell|)!}{n! (2|\ell|-n)!} (\pm i)^n
\frac{\Gamma\left(\frac{1}{2} -  |\ell|\right)}
{\left(\frac{\nu-n}{2}\right)! \,
\Gamma\left(\frac{1}{2} - |\ell| + \frac{n-\nu}{2}\right)} \,,
\end{equation}
where the $\pm$ is $+$ for $\ell \ge 0 $ and $-$ for $\ell < 0$,
and the sum starts at $n=0$ for $\nu$ even and $n=1$ for $\nu$ odd,
and then in both cases goes in steps of $2$.  For a configuration
where the displacement from the focus of the parabolic cylinder to the
center of the ordinary cylinder is parameterized in ordinary
cylindrical coordinates by distance $r_0$ and angle $\theta_0$  from
the $x$-axis, the Casimir interaction energy is then
\begin{equation}
\frac{\cal E}{\hbar c L} = 
\int_0^\infty \frac {d\kappa}{2 \pi} 
\int_{-\infty}^\infty \frac {dk_z}{2 \pi}
\log \det \left(\mathbbm{1}_{\nu \nu'} - 
\left({\cal T}_{\nu}^C\right)^{-1}
\sum_{\ell=-\infty}^\infty  
{\cal V}_{\nu \ell}(r_0, \theta_0) {\cal T}_{\ell}^{O}
\hat{\cal V}_{\nu' \ell}(r_0, \theta_0)
\right)  \,,
\label{eqn:Eparordint}
\end{equation}
where
${\cal T}_{\nu}^{C}$ is the $T$-matrix for the (outer) parabolic cylinder,
${\cal T}_{\ell}^{O}$ is the $T$-matrix for the (inner) ordinary
cylinder, and
\begin{equation}
{\cal V}_{\nu \ell}(r_0, \theta_0) = 
\hat {\cal V}_{\nu \ell}(r_0, \pi-\theta_0) =
\frac{1}{\sqrt{\nu!\sqrt{2\pi}}} \sqrt{\frac{4 (-1)^\ell}{i}}
\left[
\sum_{\ell_0=-\infty}^{\infty}
\beta_{\nu, \ell+\ell_0} e^{i\ell_0 \theta_0}
i^{\ell_0} I_{\ell_0}(\sqrt{\kappa^2+k_z^2} r_0)
 \right] \,.
\end{equation}

\begin{figure}
\hfill
\includegraphics[width=0.525\linewidth]{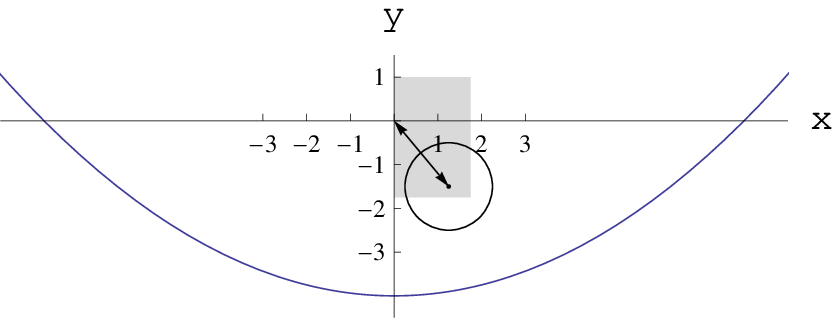}\vspace{0.125in}
\hfill
\includegraphics[width=0.45\linewidth]{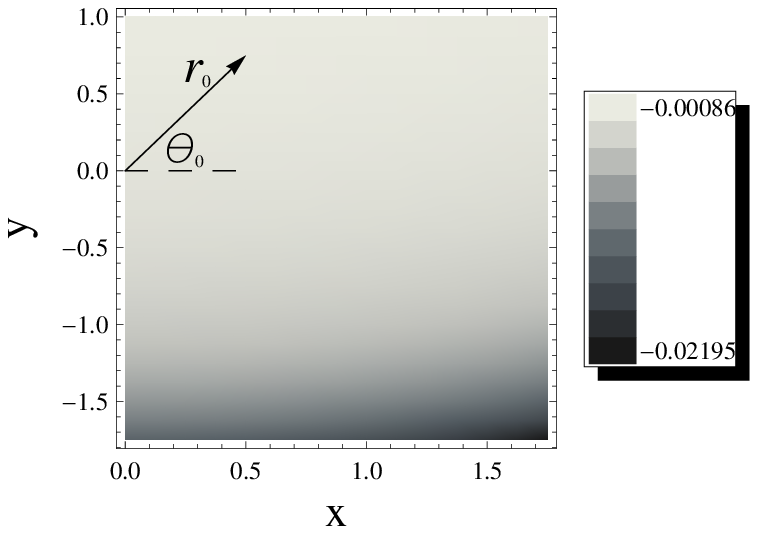}
\caption{Casimir interaction energy for an ordinary cylinder inside a
parabolic cylinder.  The left panel shows the geometry, with the
focus of the parabolic cylinder placed at the
origin.  We choose the same radii as Ref.~\cite{Lombardo}, $\mu_0 =
\sqrt{8}$ and ${\cal R}=1$, so that the vertex line of the parabolic
cylinder lies at $x=0$, $y=-4$.  The right panel shows the
Casimir interaction energy per unit length ${\cal E}/(\hbar c L)$ as a
function of $x$ and $y$, the displacement of the center of
the ordinary cylinder from the focus of the parabolic cylinder, where
the center of the ordinary cylinder lies within the shaded region of
the left panel.}
\label{fig:cylinterior}
\end{figure}

As an example, we consider a thin wire near the focal axis of a
parabolic cylinder with parabolic radius $\mu_0^2$.  To leading order in
the needle radius ${\cal R}$, we only need the $\ell=0$ Dirichlet
$T$-matrix element, which goes like $1/\log {\cal R}$ for ${\cal R}$
small.  Keeping only the leading term in $1/\log ({\cal R}/\mu_0^2)$
and also expanding in the displacement from the focus $r_0$, we obtain
\begin{equation}
\frac{\cal E}{\hbar c L} \approx
\frac{3}{32 \mu_0^4 \log \frac{{\cal R}}{\mu_0^2}}
- \frac{5}{16 \mu_0^6 \log \frac{{\cal R}}{\mu_0^2}} r_0 \sin \theta_0
+ \frac{15 }{256 \mu_0^8 \log\frac{\cal R}{\mu_0^2}} r_0^2 (9-5\cos 2\theta_0)
+ \cdots \, .
\end{equation}
Here energy is calculated in comparison
to the configuration where the ordinary cylinder is placed at $x=0$,
$y=\infty$.  In deriving this result, we have assumed that ${\cal R}
\ll \mu_0^2$ is small enough that  we can drop terms proportional to 
$\log (\mu_0^2 \sqrt{\kappa^2 + k_z^2})$ 
in comparison to terms proportional to $\log ({\cal R}/\mu_0^2)$, since for
$\sqrt{\kappa^2 + k_z^2} \gg 1/\mu_0^2$  the integrand in 
Eq.~(\ref{eqn:Eparordint}) is exponentially suppressed.
The first term gives the (negative) Casimir interaction
energy per unit length when the wire is at the focus, while the
second and third terms give the correction as it is moved a small
distance away.  The angular dependence is exactly as we would expect:  as the
wire moves closer to the vertex axis of the parabolic cylinder
($\theta_0 = -\pi/2$), the energy gets more negative; as it moves away
from the vertex axis ($\theta_0 = \pi/2$) the energy is less negative,
and if it moves in a direction perpendicular to the plane of symmetry
of the parabolic cylinder ($\theta_0 = 0$ or $\theta_0 = \pi$), the
energy is unchanged to first order.  As a result, unlike the geometric
optics calculation considered in Ref.~\cite{Ford}, here we do not see
any unusual behavior of the Casimir energy at the focus, which is in
agreement with the results in Ref.~\cite{Lombardo}.  In
Fig.~\ref{fig:cylinterior} we illustrate this result for the
case where the radius  of the ordinary cylinder and its displacement
are not small.  We choose the same radii for the parabolic and ordinary
cylinders as in Ref.~\cite{Lombardo}, and the results we obtain are
approximately in agreement with what was found there.  We cannot make
a precise comparison, however, because in that work the parabolic
cylinder is of finite size and closed at the far end.

\section{Nonzero Temperature}\label{finiteT}

It is straightforward to extend all of these results to temperature
$T\neq 0$, a subject that has been of significant recent interest
\cite{Weber}.  In each calculation, we simply 
replace the integral $\int_0^\infty
\frac{d\kappa}{2\pi}$ by the sum $\frac{T k_B }{\hbar c}
{\sum_{n=0}^\infty} '$ over Matsubara frequencies $\kappa_n = 2 \pi n
 k_B T/(\hbar c)$, where $k_B$ is Boltzmann's constant and the prime
indicates that the $n=0$ mode is counted with a weight of $1/2$
\cite{universal}.  In the classical limit of asymptotically large
temperature, only the $n=0$ term contributes.  The numerical
calculation is more cumbersome for $T\neq 0$, because for $T=0$ we
could always make the substitution
$\displaystyle \int_0^\infty \frac{d\kappa}{2\pi}
\int_{-\infty}^\infty \frac {dk_z}{2 \pi} f(\sqrt{\kappa^2 + k_z^2}) \to
\frac {1}{4 \pi} \int_0^\infty q dq f(q)$, where $q=\sqrt{\kappa^2 +
k_z^2}$, since the quantity we integrate depends only on $q$.
We find it convenient to continue to use the integration variable $q$,
since the quantity we now sum and integrate still depends only on this
quantity.  We therefore carry out the sum and integral via the
replacement
\begin{equation}
\frac{ k_B T}{\hbar c} {\sum_{n=0}^\infty}'
\int_{-\infty}^\infty \frac {dk_z}{2 \pi} f(\sqrt{\kappa_n^2+k_z^2}) \to
\frac{ k_B T}{\hbar c} \left[
\int_{0}^\infty \frac {dk_z}{2 \pi} f(k_z) 
+ \int_{0}^\infty \frac {dq}{\pi} 
\sum_{n=1}^{\lfloor \frac{\hbar c q}{2 \pi  k_B T} \rfloor}
\frac{q}{\sqrt{q^2 - \kappa_n^2}} f(q)
\right]\,,
\end{equation}
where $\lfloor x \rfloor$ denotes the greatest integer less than or
equal to $x$.

\begin{figure}[htbp]
\includegraphics[width=0.5\linewidth]{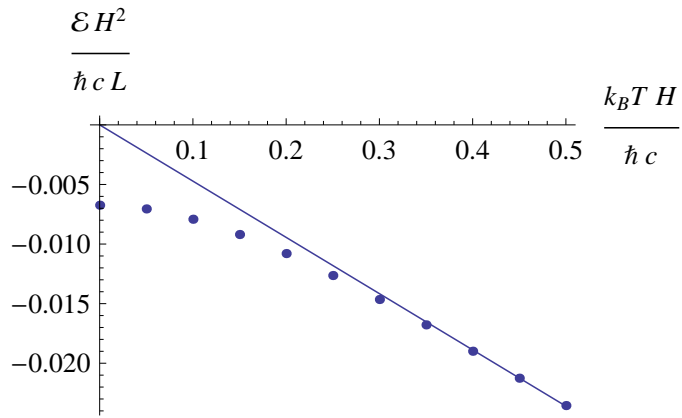}
\caption{The energy per unit length times $H^2$,
${\cal E} H^2/(\hbar c L)$, plotted versus $ k_B T H/(\hbar c)$ 
for $\theta_C=0$ and
$R=0$.  The solid line gives the $T\to \infty$ limit determined from
the lowest Matsubara frequency.  For reference, at a separation of $H=
350\ {\rm nm}$, a temperature $T=300\ {\rm K}$ corresponds to
$ k_B T H/(\hbar c) \approx 0.046$.
}
\label{fig:EvsT}
\end{figure}

As an example, we consider thermal corrections for a 
conducting half-plane (a parabolic cylinder with $R=0$) oriented
perpendicular to a conducting plane, at separation $H$.  In the 
classical limit, only the $n=0$ mode contributes and we obtain
the energy ${\cal E}/L = - k_B T C_{T=\infty}/H$, with
$C_{T=\infty} = 0.0472$.  The Dirichlet contribution to this result is
$C_{T=\infty}^D = 0.0394$, in agreement with Ref.~\cite{Gies}.  In
Fig.~\ref{fig:EvsT} the energy for this geometry is shown as a
function of temperature.  For typical separation distances at room
temperature, the thermal corrections are small.

\section{Conclusions}\label{conclude}

There are only a limited set of coordinate systems in which the vector
Helmholtz equation for electromagnetism can be solved exactly.
Taking advantage of one of these few cases, we have obtained complete
scattering amplitudes for a perfectly conducting parabolic cylinder
and employed these results to compute Casimir forces.
In principle, Casimir forces can be computed in configurations
involving parabolic  cylinders and other shapes for which scattering
amplitudes are known, as long as we can obtain the {\em
translation matrices}, which convert expressions of electromagnetic
waves between different coordinate basis, appropriate to the
individual shapes.  Following this procedure, we have computed Casimir
forces between a parabolic cylinder, a plane, an ordinary cylinder,
and a second parabolic cylinder. The formalism is versatile enough to
treat situations in which one object is enclosed in the interior of a
parabolic cylinder, and is also easily extended to finite temperatures.

We focus special attention to the limit when the radius of the
parabolic cylinder goes to zero, and it evolves into a semi-infinite
plate --- a {\em knife edge}.  In this limit we can quantify the
contribution of edges to the Casimir force.  By examining tilted
plates, we can consider a broad range of cases involving interacting
edges which should be useful to the design of microelectromechanical
devices.  Until recently, the state of art computation of Casimir
forces relied upon the PFA, which is demonstrably unreliable for a
knife edge:  A thin metal disk perpendicular to a nearby metal surface
experiences a  Casimir force described by an extension of
Eq.~(\ref{eq:zeroR}), while as indicated in Fig.~\ref{fig:EvsH}, the
PFA approximation to the energy vanishes as the thickness goes to zero. 
Based on the full result for perpendicular planes,
however, we can formulate an ``edge PFA,'' which yields the energy by
integrating $d\mathcal{E}/dL$ from Eq.~(\ref{eq:zeroR}) along the
edge of the disk.  Letting $r$ be the disk radius, in this
approximation we obtain
\begin{equation}\label{edgePFA} 
\mathcal{E}_{\hbox{\tiny Epfa}} = -\hbar c C_\perp
\int_{-r}^r (H+ r-\sqrt{r^2-x^2})^{-2} dx \xrightarrow{H/r\to0} -\hbar
c C_\perp \pi \sqrt{r/(2H^3)}\,,
\end{equation} 
which is valid if the thickness of the disk is small compared to its
separation  from the plane.  (For comparison, note that the ordinary
PFA for a metal  sphere of radius $r$ and a plate is proportional to
$r/H^2$.)

A disk may be more experimentally tractable than a plane, since its edge
does not need to be maintained parallel to the plate.  One possibility
is a metal film, evaporated onto a substrate that either has low
permittivity or can be etched away beneath the edge of the deposited
film. Micromechanical torsion oscillators, which have already been
used for Casimir experiments \cite{Decca07}, seem readily adaptable
for testing Eq.~(\ref{eq:Etheta}).  Because the overall strength of
the Casimir effect is weaker for a disk than for a sphere, observing
Casimir forces in this geometry will require greater sensitivities or
shorter separation distances than the sphere-plane case.  As the
separation gets smaller, however, the dominant contributions arise
from higher-frequency fluctuations, and deviations from the perfect
conductor limit can become important.  While the effects of finite
conductivity could be captured by an extension of our method, the
calculation becomes significantly more difficult in this case because
the matrix of scattering amplitudes is no longer diagonal.

To estimate the range of important frequencies, we
consider $R\ll H$ and $\theta_C=0$.  In this case, the integrand
in Eq.~(\ref{eq:zeroR}) is strongly peaked around $q \approx 0.3/H$.
As a result, by including only values of $q$ up to $2/H$, we still
capture $95\%$ of the full result (and by going up to $3/H$ we include
99\%).  This truncation corresponds to a minimum ``fluctuation
wavelength'' $\lambda_{\hbox{\tiny min}} = \pi H$.  For the perfect
conductor approximation to hold, $\lambda_{\hbox{\tiny min}}$ must be
large compared to the metal's plasma wavelength $\lambda_p$, so that
these fluctuations are well described by assuming perfect
reflectivity.  We also need the thickness of the disk to be small
enough compared to $H$ that the deviation from the proximity force
calculation is evident (see Fig.~\ref{fig:EvsH}), but large enough
compared to the metal's skin depth $\delta$ that the perfect conductor
approximation is valid.  For a typical metal film, $\lambda_p \approx
130\ {\rm nm}$ and $\delta \approx 25\ {\rm nm}$ at the relevant
wavelengths.  For a disk of radius $r=100\ \mu{\rm m}$, the present
experimental frontier of $0.1\ {\rm pN}$ sensitivity corresponds to a
separation distance $H\approx 350\ {\rm nm}$, which then falls within
the expected range of validity of our calculation according to these
criteria.  The force could also be enhanced by connecting several
identical but well-separated disks.  In that case, the same force could
be measured at a larger separation distance, where our calculation is
more accurate.  In the case of overlapping planes, the correction to
the traditional PFA energy is of a similar magnitude to the total
force for perpendicular planes in the above example, and thus should 
also be measurable at these separations.  
We have shown that thermal corrections are generally small at room 
temperature for typical separations, and furthermore our methods allow 
these corrections to be computed precisely.

\section{Acknowledgments}

We thank U. Mohideen for helpful discussions
and F. Khoshnoud for correspondence regarding the example of
overlapping planes.
This work was supported in part by the National Science Foundation
(NSF) through grants PHY08-55426 (NG), DMR-08-03315 (SJR and
MK),  Defense Advanced Research Projects Agency (DARPA) contract
No. S-000354 (SJR, MK, and TE), and by the U. S. Department of Energy
(DOE) under cooperative research agreement \#DF-FC02-94ER40818 (RLJ).


\end{document}